# A Non-linearized PLS Model Based on Multivariate Dominant Factor for Laser-induced Breakdown Spectroscopy Measurements

Zhe Wang, Jie Feng, Lizhi Li, Weidou Ni, Zheng Li[*]

*State Key Lab of Power Systems, Department of Thermal Engineering, Tsinghua-BP Clean Energy Center, Tsinghua University, Beijing, China*

## Abstract

A multivariate dominant factor based non-linearized PLS model is proposed. The intensities of different lines were taken to construct a multivariate dominant factor model, which describes the dominant concentration information of the measured species. In constructing such a multivariate model, non-linear transformation of multi characteristic line intensities according to the physical mechanisms of lased induced plasma spectrum were made, combined with linear-correlation-based PLS method, to model the nonlinear self-absorption and inter-element interference effects. This enables the linear PLS method to describe non-linear relationship more accurately and provides the statistics-based PLS method with physical backgrounds. Moreover, a secondary PLS is applied utilizing the whole spectra information to further correct the model results. Experiments were conducted using standard brass samples. Taylor expansion was applied to make the nonlinear transformation to describe the self-absorption effect of *Cu*. Then, line intensities of another two elements, *Pb* and *Zn*, were taken into account for inter-element interference. The proposed method shows a significant improvement when compared with conventional PLS model. Results also show that, even compared with the already-improved baseline dominant-factor-based PLS model, the present PLS model based on the multivariate dominant factor yields the same calibration quality ($R^2$=0.999) while decreasing the RMSEP from 2.33% to 1.97%. The overall RMSE was also improved to 1.05% from 1.27%.

## 1 Introduction

Laser-induced breakdown spectroscopy (LIBS) has been increasingly interested in as an analytical tool in various fields [1-10], but how to achieve accurate quantitative results remains as a bottleneck for successful commercial application. The conventional univariate method is based on the physical background of LIBS that a greater density of species in the plasma results in higher measured characteristic line intensity. However, since the measured line intensity is unavoidably influenced by many other factors, such as uncontrollable experimental parameter fluctuations and the physical and chemical matrix effect [11], the accuracy of the conventional univaritate model is deteriorated and the applicability of the model is limited.

[*] Corresponding author. Tel: +86 10 62795739; fax: +86 10 62795736
E-mail address: lz-dte@tsinghua.edu.cn

In order to compensate for deviations from different sources, many researchers now use multivariate analysis to extract more quantitative information from the whole spectra. The emerging partial least squares regression tool has shown great potential in LIBS analysis [12-18]. Generally, because PLS utilizes all spectral information, the model shows more accurate results than the conventional univariate method. However, conventional PLS is, basically, a statistical method that only considers linear correlations between the dependent and independent variables, in this case element concentrations of calibration samples and the spectra, while neglecting the physical derivation of plasma and not reconciling the fact that the relationship between the spectra often shows a non-linear behavior due to the non-linear effects such as self-absorption effect and inter-elemental interference. Therefore, linear PLS cannot guarantee accurate prediction results. For instance, Fink et al. found that the relative prediction error for element measurement in recycled thermoplasts reached the order of 15-25% using PLS [19]. The source of the error, they believe, was fluctuation in the matrix and inconstant ablation behavior. Sirven et al. found that the artificial neural network (ANN) approach achieved better quantitative results than PLS since ANN could deal with non-linear relationship [20]. How to combine PLS with the physical principles and induct non-linear factor into PLS are the potential directions to improve the accuracy of PLS.

To reduce the disadvantages of PLS, a multivariate method based on dominant factor was proposed in a previous paper [21]. That method constructed a dominant factor, which is an explicitly extracted expression for element concentration calculation and takes a dominant portion of the total model results, and further compensated for the residuals between the dominant factor model and nominal elemental concentrations with PLS. The application of this model to brass alloy samples proved that combining physical principles and inducting non-linear correlation can largely improve the accuracy of PLS.

It was found that the more accurate the dominant factor model established, the more accurate the final PLS correction model results obtained. Therefore, establishing an accurate dominant factor model and making PLS correction for the residuals needless may be the ultimate goal of the development of this method. Based on the work in Ref [21], this paper further utilizes various characteristic line intensities (including atomic and ionic lines) of different elements to construct a more comprehensive and accurate dominant factor, making the whole model more robust in improving the prediction results for the same brass alloy sample set as Ref. [21].

## 2 Methods

### 2.1 Basis

In Ref [21], based on the understanding of the evolution of the characteristic line intensity along with the elemental concentration change in LIBS measurement, a dominant factor was extracted, and PLS approach was then applied to further improve the dominant factor model. In the application to brass alloy samples for *Cu* concentration measurement, the final model expression was obtained as:

$$C_{Cu} = C_0 \ln(\frac{bC_0}{a + bC_0 - I_{Cu}}) + a_0 + a_1 I_{Pb} + a_2 I_{Pb}^2 + \sum_{j=1}^{n} b_j I_j + b_0 \quad (1)$$

In Eq. 1, $C_{Cu}$ is the model calculated element concentration, and $C_0 \ln(\frac{bC_0}{a + bC_0 - I_{Cu}})$ describes self-absorption while $a_0 + a_1 I_{Pb} + a_2 I_{Pb}^2$ models inter-element interference. $C_0$ is considered as the "saturation concentration", $I_{Cu}$ is one individual *Cu* characteristic line integrated intensity, $I_{Pb}$ is the integrated intensity of single *Pb* line and $a$, $b$, $a_0$, $a_1$, $a_2$ are the constants calculated by the best-fitting technique. The entire item $C_0 \ln(\frac{bC_0}{a + bC_0 - I_{Cu}}) + a_0 + a_1 I_{Pb} + a_2 I_{Pb}^2$ is defined as the dominant factor. In addition, the residuals (the differences between the calculation results of dominant factor and the real elemental concentration) were compensated by PLS with full-spectra input, as expressed by $\sum_{j=1}^{n} b_j I_j + b_0$ in Eq. 1, where $I_j$ is the spectral intensities at different wavelength, while, $b_0$ and $b_j$ ($j$=1, 2, ..., n; $n$ is the number of spectral points used in PLS regression) are the regression coefficient. The detailed process to obtain the dominant factor model, which is represented by Eq. 1, can be referred to Ref. [21].

In essence, the dominant factor in the Ref. [21] model can be regarded as the first principal component of the conventional PLS model. That is, based on the physical laws, the model used the dominant factor to extract the principal component manually and explicitly, and non-linear factor was inducted into such a principal component. Therefore, this model partially overcame the disadvantages of PLS, such as the lack of physical background and consideration of only linear correlation. Results in Ref. [21] showed that the model based on the dominant factor performed very well for LIBS measurement of *Cu* in brass alloys. Compared with conventional PLS, the Ref. [21] model maintained the calibration accuracy and improved prediction results by

56%.

However, in Ref. [21], the extraction of the dominant factor was limited in that it merely utilized only a single *Cu* atomic characteristic line to consider self-absorption in addition to the arduous dominant factor extraction process. Moreover, in modeling inter-element interference, only the individual characteristic line intensity of another element which demonstrated the most significant linear correlation was applied. Because of the complexity of self-absorption and inter-element interference as well as the uncertainty of concentration information contained in a single line intensity, the utilization of a single characteristic line for each element to model self-absorption and inter-element interference may still contain large uncertainty and inaccuracy to the final results. Given these shortcomings of the previous model, there is still some space to improve the accuracy of the dominant factor as well as the whole model.

## 2.2 Model descriptions

In the LIBS spectrum, there is usually more than one atomic or ionic line emitted by the same element. All these characteristic lines contain information of the elemental concentration and the interaction with other elements in the plasma. Furthermore, each characteristic line responds differently to experimental fluctuations, and correlations exist among these intensity changes and the fluctuations. As such, the use of multiple characteristic line intensities of the element of interest and other main elements to model self-absorption and inter-element interference offers the potential to obtain more comprehensive and robust results. Since these line intensities are all correlated to the concentration of the element of interest, multivariate PLS approach is then chosen.

However, PLS is a basically linear correlation multivariate method, while the dominant factor model in Ref. [21] used non-linear relationships. Therefore, non-linearity consideration should be inducted into PLS, especially when the region of interest is large. In this case, a natural way is to non-linearize the spectral intensity variables. This is simple and does not complicate the model compared to other ways of handling non-linearities [22-23]. Here is an example to illustrate how non-linear transformation helps to compensate for the non-linear dependence. For the non-linear quadratic function $Y=aX^2+b$, there is a quadratic dependence of $Y$ on $X$. If $X$ is transformed non-linearly to $X^2$ as a new independent variable $X'=X^2$, then linear function $Y=aX'+b$ can be used to describe the nonlinear relation between $Y$ and $X$. Therefore, after such non-linear transformation, linear PLS is more capable of modeling non-linearity between corrected independent and dependant variables.

In the present approach, the non-linear transformation is applied to compensate for the non-linear spectral intensity dependence on concentration, which helps to construct a more advanced and effective dominant factor model using PLS. Based on the multivariate dominant factor, PLS is further employed to correct the residual of the dominant factor with information from the full spectra.

It should be noted that inappropriate non-linear transformation may yield even less accurate measurement results in that inappropriate non-linearization might further distort the correlation, resulting less capability in reflecting the should-be nonlinear relationship. Therefore, in order to ensure appropriate non-linear transformation, the physical laws of plasma spectroscopy were carefully applied to conduct the non-linear transformation of line intensities. Based on the dominant factor model in Ref. [21], the present multivariate dominant factor model is established as follows.

Firstly, the item $C_0 \ln(\frac{bC_0}{a+bC_0-I_{Cu}})$ for self-absorption in Eq. 1 requires arduous calculations to determine the constants $a$ and $b$ when multiple characteristic lines are considered, since these constants change with the applied $I_{Cu}$ at different wavelength. To find a more simplified expression and reduce calculation complexity, Taylor expansion was used. In addition, it was found that after the normalization of spectral data, the values of all $Cu$ line intensities are on the magnitude of $10^{-4}$, being very close to zero. Thus, $C_0 \ln(\frac{bC_0}{a+bC_0-I_{Cu}})$ can be expanded at $I_{Cu}=0$ using the Taylor series as follows:

$$C_0 \ln(\frac{bC_0}{a+bC_0-I_{Cu}}) = C_0 \ln(\frac{bC_0}{a+bC_0}) + \frac{C_0}{a+bC_0} I_{Cu} + \frac{C_0}{2(a+bC_0)^2} I_{Cu}^2$$
$$+ \frac{C_0}{6(a+bC_0)^3} I_{Cu}^3 + ... \qquad (2)$$

Standard practice in the industry express calibration curves as third-order polynomials (over the whole calibration range) [23], so the present approach applies the forms of $I_{Cu}$, $I_{Cu}^2$ and $I_{Cu}^3$ to transform all $Cu$ characteristic line intensities as new independent variables. That is, in applying PLS with the new input variables, the self-absorption effect was modeled with the combination of the original, squares, and cubes of all $Cu$ characteristic line intensities and the self-absorption model in the dominant factor is shown as:

$$C_{Cu,SA} = \sum_{i=1}^{m} c_i I_{Cu,i} + \sum_{i=1}^{m} d_i I_{Cu,i}^2 + \sum_{i=1}^{m} e_i I_{Cu,i}^3 \qquad (3)$$

where $C_{Cu,SA}$ is the calculated elemental concentration of the dominant factor considering self-absorption, $I_{Cu,i}$ is the integrated intensity of $Cu$ lines at different wavelengths, and $c_i$, $d_i$ and $e_i$ ($i=1$, 2, …, $m$; $m$ is the number of $Cu$ lines) are coefficients. Considering the co-linearity among the independent variables, the powerful PLS is a good method to calculate the coefficients in Eq. 2.

Secondly, inter-element interference should also be considered in the dominant factor. In Ref. [21], the quadratic curve could not perfectly describe the correlation between a single *Pb* line intensity and the residuals after self-absorption modeling. The reason might be that more line information correlated to inter-element interference is needed to be taken into account to improve the accuracy of inter-element interference modeling. Therefore, in this model, more *Pb* lines were included in the dominant factor, and since the mechanism of inter-element interference is very complicated and remains unclear, PLS is also applied to model the phenomenon implicitly. Meanwhile, to avoid inter-element interference influence that is muted by other fluctuations and thus cannot be accurately compensated for during PLS calculation, this model tries to adopt new variables consisting of various *Pb* line intensities multiplied and divided by a *Cu* line intensity which shows most correlation with *Cu* concentration, making the new variables more sensitive to *Cu* concentration change, and thus calculation results by PLS more robust. The *Cu* line at 570.024 nm is chosen, as it provides the best fitting result in univariate modeling. The inter-element interference is modeled as follows:

$$C_{Cu,\text{Interf}} = \sum_{s=1}^{p} f_s I_{Pb,s} I_{Cu,\text{SCL}} + \sum_{s=1}^{p} g_s \frac{I_{Pb,s}}{I_{Cu,\text{SCL}}} + \sum_{s=1}^{p} h_s I_{Pb,s}^2 I_{Cu,\text{SCL}} + \sum_{s=1}^{p} k_s \frac{I_{Pb,s}^2}{I_{Cu,\text{SCL}}} \tag{4}$$

where $C_{Cu,\text{Interf}}$ is the calculated result from inter-element interference modeling, and $I_{Pb,s}$ are *Pb* line intensities at different wavelengths, respectively. $I_{Cu,\text{SCL}}$ is the integrated intensity of specific characteristic *Cu* line at 570.024 nm. $f_s$, $g_s$, $h_s$, $k_s$ (*s*=1，2，…，*p*；*p* is the number of *Pb* lines) are coefficients calculated by PLS regression. As described above, the selection of the products and quotients of the *Pb* line intensities and a specific *Cu* emission line was a try-out to model the complicated inter-elemental interference effects. As the understanding to inter-elemental interference furthers, there may be more suitable transformations found to improve the final results.

Then the multivariate dominant factor considering self-absorption and inter-element interference is expressed as:

$$C'_{Cu} = C_{Cu,\text{SA}} + C_{Cu,\text{Interf}} + b_{Cu} \tag{5}$$

where $C'_{Cu}$ is the calculated elemental concentration of the dominant factor considering self-absorption and inter-element interference, and $b_{Cu}$ is the coefficient decided by PLS regression.

In addition, for other samples, it may need to take steps as shown in Ref. [21] to obtain a similar formula as Eq. 1. That is, the Eq. 1 in the present work cannot be directly applied to other types of samples. The nonlinear transformation will therefore be different accordingly to maintain the essence of transforming the raw spectral data with physical mechanism.

Finally, the residuals of the dominant factor, after modeling self-absorption and inter-element interference, mainly come from other unknown or incalculable fluctuations which are difficult to accurately compensate for. Such fluctuations are somehow correlated to the full spectral information and thus can be implicitly modeled by PLS with full spectra input. Just as in Eq. 1, residual correction is added to the equation so that the final expression of the model is:

$$C_{Cu} = C'_{Cu} + \sum_{j=1}^{n} b_j I_j + b_0 \qquad (6)$$

Compared with the approach in Ref. [21], the advantages of the PLS model based on a multivariate dominant factor are that: 1) the dominant factor employs information from multiple characteristic lines into consideration to construct a comprehensive index to reflect the elemental concentration, being capable of modeling self-absorption and inter-element interference with more complete information from the spectra, and 2) the line intensities are transformed non-linearly according to physical laws before PLS regression, enabling PLS to better describe the non-linear relationship between the spectra and elemental concentration. In theory, the application of intensities after non-linear transformation improves the dominant factor accuracy, making the model obtain robust results in a wide concentration range.

## 3 Experiment set-up

Fourteen standard brass alloy samples used in the experiment are from Central Iron and Steel Research Institute of China. The main elemental concentrations are listed in **Table 1**.

Table 1. Main elemental concentrations of brass alloy samples

A Spectrolaser 4000 (XRF, Australia) was used in the present study. The instrument and configuration are described in a previous paper [24]. The laser has a wavelength of 532 nm and a pulse width of 5 ns. The laser pulse frequency is 1 Hz, and the gate time was fixed at 1 ms. To produce spectra with a high signal-to-noise ratio and keep the detector out of intensity saturation, the laser energy was optimized to be 90 mJ/pulse and the delay time was set for 2.25 μs. The spectrometers cover the spectral range from 190 to 940 nm with interval about 0.09 nm. The sample surface was cleaned by ethanol to remove contaminants before analysis and was placed on an auto-controlled X-Y translation stage exposed to open air. Before analysis, a laser pulse of 150 mJ was used to burn off contaminants. For each sample, an averaged spectrum of 35 replications at different locations on the sample surface was used to reduce the influence from sample heterogeneity and other fluctuations

The background signal was recorded by the instrument with a low-energy laser pulse and long delay time. Background subtraction was applied to partially cancel out the errors from instrumental and environmental noise. Moreover, each detected spectrum was corrected for the efficiency of the detection system, minimizing line intensity distortion from the wavelength-dependent efficiency of optics and lenses. To reduce unintended fluctuations, all the spectra were normalized to the entire spectral area, which means all raw spectra data were divided by the whole spectral area before calibration model construction.

## 4 Results and discussion

The PLS method based on a multivariate dominant factor was evaluated in terms of *Cu* concentration measurement and compared with the dominant-factor-based PLS method in Ref. [21]. The software used to calculate PLS was SPSS 17.0 (SPSS Inc., Chicago, Illinois, USA). The coefficient of determination ($R^2$) was used to evaluate the model calibration quality, and the root mean square error of prediction (RMSEP) was applied to assess the prediction results. $R^2$ and RMSEP of an accurate model should be close to 1 and 0, respectively. Furthermore, the root mean square error of both the calibration and prediction samples (RMSE) was taken to determine the overall quality of the model. A better model has a smaller value of RMSE. Ten samples were used to construct the calibration model, while ZBY906, ZBY907, ZBY924 and ZBY927 were selected to estimate the measurement prediction. The prediction samples were picked because the *Cu* concentrations of ZBY907 and ZBY924 are inside the calibration sample set while those of ZBY906 and ZBY927 are out of the calibration range. Note that because *Cu* concentration only partially represents the matrix of the samples, the prediction samples might not completely stand for sample matrix in and out of the calibration range.

### 4.1 Baseline

In Ref. [21], during the approach to construct the dominant factor, a single *Cu* characteristic line was used to compensate for self-absorption based on the empirical expression. Then inter-element interference was described with a quadratic relationship considering the correlation between the residuals and line intensity of other main elements. After the dominant factor extraction, PLS was further applied to correct the residuals of the dominant factor with full spectra information. The above dominant-factor-based PLS method in Ref. [21] was chosen to be the baseline in this paper, and the calibration and prediction results are shown in **Fig. 1**. Additionally, to clearly compare the results between our present model and other models, the calibration, prediction, and overall results of dominant factor model, conventional PLS model, and dominant-factor-based PLS model are listed below in Table 2 and 3, respectively.

**Figure 1. Dominant-factor-based PLS method results in Ref. [21]**

As shown in **Fig. 1**, the absolute relative errors of prediction are still large for some samples, such as sample 927, showing that the dominant-factor-based PLS method in Ref. [21] can be further improved. Applying multiple characteristic lines to utilize more spectral information and construct a multivariate dominant factor is certainly such a potential way.

### 4.2 Multivariate dominant factor results

According to the method described in section 2, the *Cu* atomic and ionic lines and the characteristic lines of the other main element were extracted to construct the multivariate dominant factor with non-linearized PLS approach. *Pb* lines were firstly considered for inter-element interference modeling according to Eq. 1. It was found that the application of *Pb* lines alone could not perfectly compensate for inter-element interference. Considering the complexity of the inter-elemental interference, other element species in the plasma might also interfere with *Cu* line intensities, efforts were taken to include lines of other elements in the multivariate dominant factor, making it more effectively model inter-element interference. Results showed that lines of another main element, *Zn*, helped to improve the result of inter-element interference modeling, then the expression for inter-element interference modeling is as follows:

$$C'_{Cu,\text{Interf}} = C_{Cu,\text{Interf}} + \sum_{q=1}^{r} t_q I_{Zn,q} I_{Cu,\text{SCL}} + \sum_{q=1}^{r} o_q \frac{I_{Zn,q}}{I_{Cu,\text{SCL}}} \qquad (7)$$

where $I_{Zn,q}$ is the integrated intensity of *Zn* lines at different wavelengths, while $t_q$ and $o_q$ ($q=1, 2, \ldots, r$; $r$ is the number of *Zn* lines) are coefficients obtained through PLS regression. Then the final expression of the dominant factor is rewritten as:

$$C''_{Cu} = C_{Cu,\text{SA}} + C'_{Cu,\text{Interf}} + b_{Cu} \qquad (8)$$

where $C''_{Cu}$ is the calculated elemental concentration of the present dominant factor considering self-absorption and inter-element interference from *Pb* and *Zn*. The atomic and ionic lines with high signal-to-noise ratio and free of overlap, as listed in **Table 2**, were used to calculate the dominant factor.

**Table 2. Measured emission lines of *Cu*, *Pb* and *Zn***

The number of principal components was chosen to be three for the smallest RMSEP of the prediction sample set. **Figure 2** shows the calibration and prediction results of the multivariate dominant factor with non-linear transformation.

**Figure 2. Multivariate dominant factor results with non-linear transformation**

The results of the different dominant factors and conventional PLS are listed in **Table 3**. The dominant factors both achieve much better prediction results than conventional PLS, proving the improvement of a combination of physical background with PLS. Considering the two different dominant factors, the multivariate dominant factor in the present work largely improves the calibration and prediction quality compared with the dominant factor in Ref. [21]. $R^2$ is improved to 0.973 and RMSEP is decreased to 1.11% from 0.921 and 1.81%, respectively. The RMSE is lowered to 1.59% compared to 2.72% for the dominant factor in Ref. [21]. It is shown that the multivariate dominant factor in the present work utilized multiple lines to construct a more accurate indicator than the dominant factor in Ref. [21]. That is, the new dominant factor is capable of describing self-absorption and inter-element interference more comprehensively and effectively. Moreover, the multivariate dominant factor is extracted based on the line intensities after non-linear transformation, enabling PLS to model the non-linear relationship more accurately. Therefore, the multivariate dominant factor can be more accurate or robust over a wider concentration range, as demonstrated in **Table 3**. For instance, using the dominant factor in Ref. [21], the absolute relative error of prediction for sample 927 (highest *Cu* concentration out of the calibration set) was as high as 3.67%, while the value in the multivariate dominant factor was only 0.26%.

**Table 3. List of conventional PLS and different dominant factors**

It should be noted that the dominant factor in Ref. [21] was extracted through many steps of regression, while the multivariate dominant factor of the present model can be constructed with the application of PLS in a single step. This feature largely increases the self-adaption speed of the calibration model when more calibration samples are added. Additionally, with the development of understanding the mechanism of self-absorption and inter-element interference, more accurate non-linear transformation will be applicable, making PLS more capable of modeling the non-linear relationships between the spectra and concentration.

Furthermore, inappropriate non-linear transformation may result in less accurate results. For example, if the self-absorption laws of plasma are neglected and $I_{Cu,i}^4$ were randomly applied as corrected variables to describe self-absorption, the non-theoretical multivariate dominant factor is then expressed as follows:

$$C_{Cu}^{'''} = \sum_{i=1}^{m} r_i I_{Cu,i}^4 + C_{Cu,\text{interf}}^{'} + b_{Cu} \tag{9}$$

where $C_{Cu}^{'''}$ is the calculated elemental concentration of the dominant factor using inappropriate non-linear transformation for self-absorption, and $r_i$ ($i=1$, 2, …, $m$;

$m$ is the number of $Cu$ lines) is the coefficient provided by PLS regression. The results of such a multivariate dominant factor using non-linear transformation without solid physical background are shown in **Fig. 3**. The number of principal components was also set to be three, which was the same as the multivariate dominant factor based on physical principles shown in **Fig. 2**. Compared with the multivariate dominant factor with non-linear transformation based on physical principles (**Fig. 2**), the dominant factor with only quadruplicate correction of $Cu$ line intensities shown in **Fig. 3** has a lower value of $R^2$, and the RMSEP increased to 1.76% because of the inaccurate non-linear transformation, showing that such a random and inaccurate transformation might distort the relationship and impair the effect of PLS regression. This result shows that selecting a suitable non-linear transformation is one of the key steps in modeling non-linear relationships.

**Figure 3. Results for the multivariate dominant factor with random and inaccurate non-linear transformation**

Even in the multivariate dominant factor with non-linear transformation based on physical information, $R^2$ is not very close to 1 at 0.971. This residual mainly is a product of the inaccuracy of dominant factor modeling itself and many other unknown fluctuations. Considering the possible correlation between these fluctuations and various emission lines, normal PLS with full spectral input is used to further correct the residual in the following discussion.

## 4.3 PLS correction based on the multivariate dominant factor

Since it is difficult to explicitly determine the relationship between the residuals and the spectra, PLS is a good candidate to extract necessary information through statistical correlation. Background signals which are less related to the concentration information were eliminated from the input of PLS in order to avoid the negative influence from the excess of abundant noise [16, 18, 25]. The number of principal components was chosen as five to obtain the smallest RMSEP. The final calibration and prediction results are shown in **Fig. 4**.

**Figure 4. Final results of PLS correction based on the multivariate dominant factor**

**Table 4** lists the final results of the model in Ref. [21] and the present work. Compared with the multivariate dominant factor in **Table 3**, PLS correction improved the value of $R^2$ to 0.999, proving that PLS correction effectively compensated for the deviations from other fluctuations. The RMSEP of the final model (1.97%) was larger than that of the dominant factor alone (1.11%) and may be the result of a noise overfitting problem since PLS tried to correct any deviation using linear correlation and the full spectra. However, it cannot be denied that PLS correction improved the overall model quality given the smaller RMSE.

Furthermore, compared with the dominant-factor-based PLS method in Ref. [21], PLS correction based on multivariate dominant factor obtained better prediction results while maintaining the same quality of calibration. The RMSEP decreased from 2.33% in Ref. [21] to 1.97% in the present model, a reduction of almost 15% of the RMSEP in Ref. [21]. Note that the PLS model based on the multivariate dominant factor is especially effective for samples outside of the calibration range. Samples 906 (with lowest $Cu$ concentration) and 927 (with highest $Cu$ concentration), for example, had absolute relative errors of 0.91% and 0.49%, respectively, while these values were 2.35% and 3.30% in the dominant-factor-based PLS method in Ref. [21]. Also, the improvement of the dominant factor resulted in a final model of higher quality, confirming that the more accurate the dominant factor model constructed, the more accurate the final model results obtained.

**Table 4. Summary of the different model final results**

It should be emphasized that PLS correction must be carefully chosen according to the different situations. If the calibration samples are limited and the multivariate dominant factor is not accurate enough, meaning more information in the spectra is required to model the fluctuations and construct a robust model, PLS is needed to correct the residuals and thus improve the final results. If with the development of dominant factor modeling and an increase in the number of calibration samples, the multivariate dominant factor itself will be able to extract enough useful information in the spectra to accurately model various fluctuations using statistical correlation. Under such circumstance, the secondary PLS for residue correction may be unnecessary. For the present case, overall speaking, the secondary PLS still improves the model results and is complementary to the multivariate dominant factor model.

## 5 Conclusion

A non-linearized PLS model based on multivariate dominant factor for determination of Cu concentration in brass alloys samples is presented. The multivariate model utilizes multiple spectral information to address issues of self-absorption, inter-element interference with nonlinear transformation in combination with PLS approach. A secondary PLS model is thereafter applied to further compensate for residual errors of the multivariate dominant factor model. This approach enables the linear correlation based PLS method to deal with nonlinear relationship between the spectral intensities and interested species concentration. Furthermore, due to the nonlinear transformation was applied according to plasma spectrum physical mechanism, the present approach provides the statistical PLS approach with physics background to some extent.

Based on the empirical expression of self-absorption and Taylor expansion, $I_{Cu}$,

$I_{Cu}^2$, $I_{Cu}^3$ were taken as corrected independent variables in describing self-absorption.

For inter-element interference, line intensities of *Pb* and *Zn* were also considered to establish a multivariate dominant factor to describe inter-element interference. The inclusion of additional characteristic lines leaded to significant improvement over the conventional PLS method and already-improved dominant-factor-based PLS method in Ref. [21]. Basically, three different application methods of PLS keep the same calibration quality ($R^2$=0.999), while the nonlinearirzed PLS model based on the multivariate dominant factor results in a much improved RMSEP and RMSE. Compared with the already-improved dominant-factor based PLS model, the RESEP was improved from 2.33% to 1.97% and the overall RMSE was also decreased to 1.05% from 1.27%. Note that the RMSEP and RMSE for conventional PLS were much larger (5.25% and 2.81%, respectively).

## Acknowledgement

The authors acknowledge the financial support from the Chinese governmental "863" project (NO. 20091860346) and "973" project (NO. 2010CB227006)

## References

[1]. J.L. Gottfried, F.C. De Lucia, C.A. Munson, A.W. Miziolek, Laser-induced breakdown spectroscopy for detection of explosives residues: a review of recent advances, challenges, and future prospects, *Anal Bioanal Chem* 2009, **395**, 283-300.

[2]. D.C. Alvey, K. Morton, R.S. Harmon, J.L. Gottfried, J.J. Remus, L.M. Collins, M.A. Wise, Laser-induced breakdown spectroscopy-based geochemical fingerprinting for the rapid analysis and discrimination of minerals: the example of garnet, *Appl Optics*, 2010, **49**, 168-180.

[3]. F.C. Alvira, F.R. Rozzi, G.M. Bilmes, Laser-induced breakdown spectroscopy microanalysis of trace elements in homo sapiens teeth, *Appl Spectrosc*, 2010, **64**, 313-319.

[4]. J.M. Andrade, G. Cristoforetti, S. Legnaioli, G. Lorenzetti, V. Palleschi, A.A. Shaltout, Classical univariate calibration and partial least squares for quantitative analysis of brass samples by laser-induced breakdown spectroscopy, *Spectrochim Acta B*, 2010, **65**, 658-663.

[5]. J. Anzano, B. Bonilla, B. Montull-Ibor, R. Lasheras, J. Casas-Gonzalez, Classifications of plastic polymers based on spectral data analysis with laser induced breakdown spectroscopy, *J Polym Eng*, 2010, **30**, 177-187.

[6]. M.E. Asgill, M.S. Brown, K. Frische, W.M. Roquemore, D.W. Hahn, Double-pulse and single-pulse laser-induced breakdown spectroscopy for distinguishing between gaseous and particulate phase analytes, *Appl Optics*, 2010, **49**, 110-119.

[7]. O. Balachninaite, A. Baskevicius, K. Stankeviciute, K. Kurselis, V. Sirutkaitis, Double-pulse laser-induced breakdown spectroscopy with 1030 and 257.5 nm wavelength femtosecond laser pulses, *Lith J Phys*, 2010, **50**, 105-110.

[8]. M. Baudelet, C.C.C. Willis, L. Shah, M. Richardson, Laser-induced breakdown spectroscopy of copper with a 2 mu m thulium fiber laser, *Opt Express*, 2010, **18**, 7905-7910.

[9]. S. Beldjilali, D. Borivent, L. Mercadier, E. Mothe, G. Clair, J. Hermann, Evaluation of minor

element concentrations in potatoes using laser-induced breakdown spectroscopy, *Spectrochim Acta B*, 2010, **65**, 727-733.

[10]. A.A. Bol'shakov, J.H. Yoo, C.Y. Liu, J.R. Plumer, R.E. Russo, Laser-induced breakdown spectroscopy in industrial and security applications, *Appl Optics*, 2010, **49**, 132-142.

[11]. S.M. Clegg, E. Sklute, M.D. Dyar, J.E. Barefield, R.C. Wiens, Multivariate analysis of remote laser-induced breakdown spectroscopy spectra using partial least squares, principal component analysis, and related techniques, *Spectrochim Acta Part B*, 2009, **64**, 79-88.

[12]. F.B. Gonzaga, C. Pasquini, A complementary metal oxide semiconductor sensor array based detection system for laser induced breakdown spectroscopy: Evaluation of calibration strategies and application for manganese determination in steel, *Spectrochim Acta B*, 2008, **63**, 56-63.

[13]. M.Z. Martina, N. Labbe, T.G. Rials, S.D. Wullschleger, Analysis of preservative-treated wood by multivariate analysis of laser-induced breakdown spectroscopy spectra, *Spectrochim Acta B*, 2005, **60**, 1179-1185.

[14]. J. L. Luque-Garcí­a, R. Soto-Ayala, M.D.L. de Castro, Determination of the major elements in homogeneous and heterogeneous samples by tandem laser-induced breakdown spectroscopy-partial least square regression, *Microchem J*, 2002, **73**, 355-362.

[15]. N. Labbe, I.M. Swamidoss, N. Andre, M.Z. Martin, T.M. Young, T.G. Rials, Extraction of information from laser-induced breakdown spectroscopy spectral data by multivariate analysis, *Appl Optics*, 2008, **47**, 158-165.

[16]. M.C. Ortiz, L. Sarabia, A. Jurado-López, M.D.L. de Castro, Minimum value assured by a method to determine gold in alloys by using laser-induced breakdown spectroscopy and partial least-squares calibration model, *Anal Chim Acta*, 2004, **515**, 151-157.

[17]. M.M. Tripathi, K.E. Eseller, F.Y. Yueh, J.P. Singh, Multivariate calibration of spectra obtained by Laser Induced Breakdown Spectroscopy of plutonium oxide surrogate residues, *Spectrochim Acta B* 64, 2009, **64**, 1212-1218.

[18]. A. Jurado-López, M.D.L. de Castro, Rank correlation of laser-induced breakdown spectroscopic data for the identification of alloys used in jewelry manufacture, *Spectrochim Acta B*, 2003, **58**, 1291-1299.

[19]. H. Fink, U. Panne, R. Niessner, Process analysis of recycled thermoplasts from consumer electronics by laser-induced plasma spectroscopy, *Anal Chem*, 2002, **74**, 4334-4342.

[20]. J.B. Sirven, B. Bousquet, L. Canioni, L. Sarger, Laser-induced breakdown spectroscopy of composite samples: Comparison of advanced chemometrics methods, *Anal Chem*, 2006, **78**, 1462-1469.

[21]. Z. Wang, J. Feng, L. Li, W. Ni, Z. Li, A multivariate model based on dominant factor for laser-induced breakdown spectroscopy measurements, *J Anal Atom Spectrom*, 2011, DOI:10.1039/C1JA10041F.

[22]. A. Berglund, S. Wold, INLR, Implicit non-linear latent variable regression, *J Chemometr*, 1997, **11**, 141-156.

[23]. F.R. Doucet, T.F. Belliveau, J.L. Fortier, J. Hubert, Use of chemometrics and laser-induced breakdown spectroscopy for quantitative analysis of major and minor elements in aluminum alloys, *Appl Spectrosc*, 2007, **61**, 327-332.

[24]. J. Feng, Z. Wang, Z. Li, W. Ni, Study to reduce laser-induced breakdown spectroscopy measurement uncertainty using plasma characteristic parameters, *Spectrochim Acta B*, 2010, **65**, 549-556.

[25]. J. Amador-Hernandez, L.E. García-Ayuso, J.M. Fernández-Romero, M.D.L. de Castro, Partial least squares regression for problem solving in precious metal analysis by laser induced breakdown spectrometry, *J Anal At Spectrom*, 2000, **15**, 587-593.